\documentclass[pra]{revtex4}
\usepackage{graphicx}

\begin{document}

\title{Running-phase state in a Josephson washboard potential}

\author{G.~S.~Paraoanu}
\affiliation{NanoScience Center and Department of Physics, University of Jyv\"askyl\"a,
P.O.~Box 35 (YFL), FIN-40014 University of Jyv\"askyl\"a, FINLAND}

\begin{abstract}

We investigate the dynamics of the phase variable of an ideal underdamped Josephson junction in switching 
current 
experiments. These experiments have provided the first evidence for macroscopic quantum tunneling in large 
Josephson junctions and are currently used for state read-out of superconducting qubits.
We calculate the shape of the resulting macroscopic wavepacket and find that
the propagation of the wavepacket long enough after a switching event leads to an average 
voltage increasing linearly with time.

\end{abstract}

\maketitle


The dynamics of a large underdamped Josephson junction characterized by a 
capacitance $C$ and Josephson energy $E_{J}$ can be described by the motion of a particle in a 
washboard potential $U(\gamma )= E_{J}(1 - \cos\gamma ) + I\bar\Phi_0 \gamma $. 
The particle has $C$ as the mass, the flux $\bar{\Phi}_{0}\gamma$ as the coordinate
and the charge on the capacitor $Q$ as the canonically conjugate momentum. Here $\gamma$ is the
phase difference of the superconducting order parameter across the junction and $\bar \Phi_{0} = 
\Phi_{0} /2\pi = \hbar/2e$ is the flux quanta divided by $2\pi$. Much attention has been given since the 
discovery of the Josephson effect to the switching dynamics of the junction in the thermal activation 
regime and in the macroscopic tunneling (MQT) regime. Surprisingly, while the description of the state
of the junction before a switching event and calculations of the corresponding probability has been a topical issue
for many decades, what happens with the quantum state of the junction after tunneling did not receive that 
much attention. It is argued \cite{tinkham,switch} that the junction ends up in a running-phase state, 
with the voltage $Q/C$ increasing until it becomes sufficiently large so that the transport could be 
done through quasiparticle excitations. 
However, a quantum mechanical description of this state is missing. Much of what we understand about 
the running-wave state, for instance
the physics of the retrapping current, comes from assuming a quasi-classical dynamics.

\begin{figure}[htb]
\includegraphics[width=84truemm]{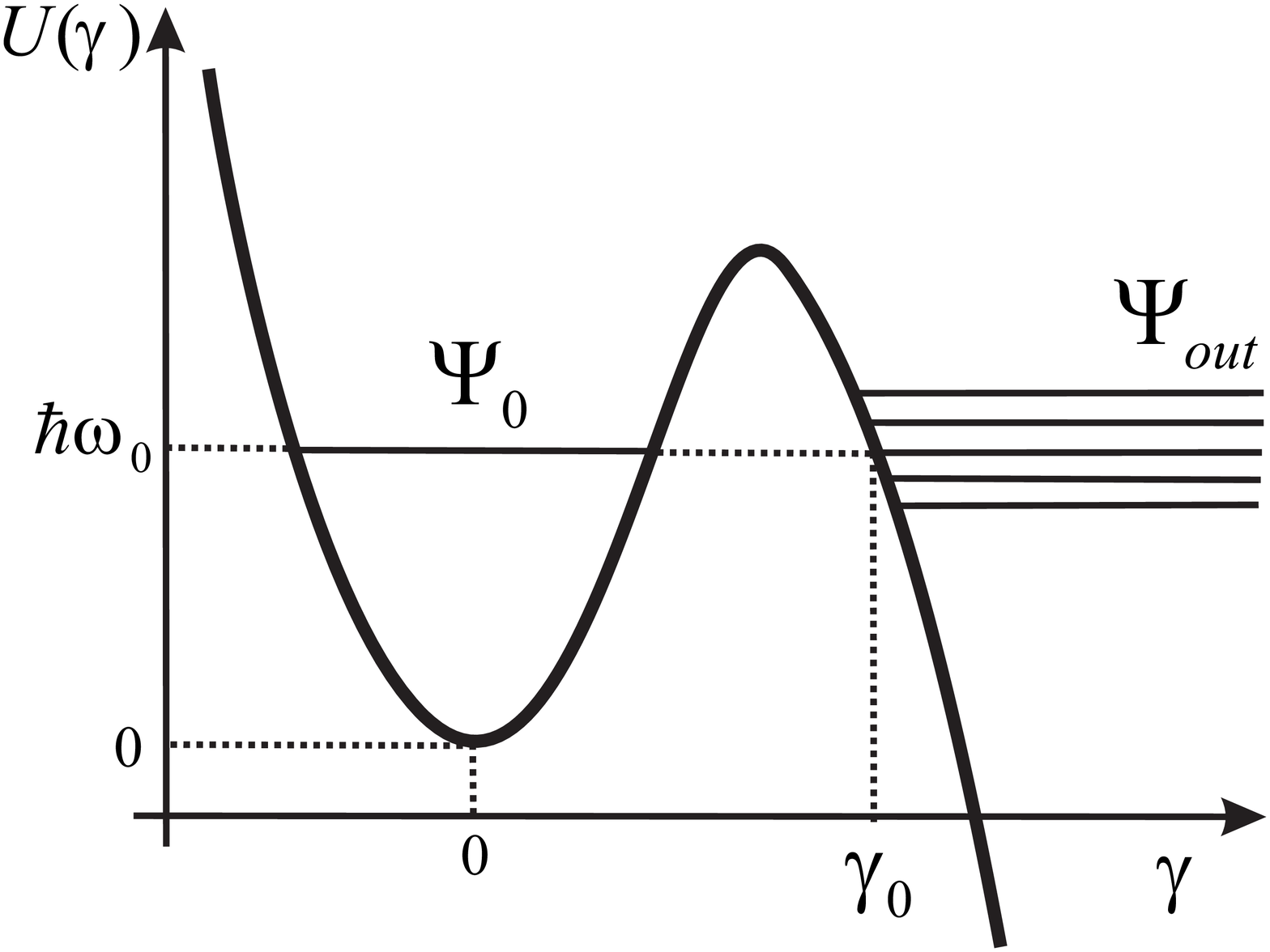}
\caption{Tunneling out of one of the metastable wells of the washboard potential. }
\label{tfig}
\end{figure}

In this paper we give an explicit formula for the macroscopic wavefunction of an ideally underdamped
 junction after a MQT switching event. 
If the switching probability is exponential, 
which is the case for all the theoretical models and also confirmed experimentally,
one expects \cite{suppl} the following expression for the dynamics of the wavefunction
\begin{equation}
|\Psi (t)\rangle = e^{-\Gamma t /2}e^{-i\omega_0 t}|\Psi_{0}\rangle + |\Psi_{out}(t)\rangle.
\label{psii}
\end{equation} 
In this equation, $\Psi_{0}$ is the initial state, coresponding to a bound state inside one of the 
metastable wells, while $\Psi_{out}(t)$ is the wavefunction of the particle corresponding to states in the 
continuum, outside the well (Fig. \ref{tfig}). This expression gives indeed an exponentially decreasing 
probability for
the particle to be inside the well, with lifetime $\Gamma ^{-1}$.
In the following, we are interested in the structure of $\Psi_{out}(t)$. 

To solve this problem, the standard approach is to start with a wavepacket localized initially in one of the 
metastable wells, and then expand it and evolve it in the eigenfunctions of the full Hamiltonian.
This procedure works for simple potentials \cite{deltapotential}, but even in these cases the solutions are
complicated. Fortunately, unlike problems in scattering theory, in condensed
matter the frequent situation is that we do not need an exact solution of the Scr\"odinger problem for tunneling,
but rather we are interested in the most generic features of it. In most cases in solid state physics, 
tunneling is simply
treated as a process that annihilates a particle on some mode of a solid and creates one on another mode.
We will approach our problem in the same spirit \cite{cond}.
A good approximation in MQT is that no other state within the well is involved
with the exception of the state with energy $\omega_0$ in which the system is prepared, $|\Psi_0 \rangle$; 
therefore one can write a reduced Hamiltonian of the form
\begin{equation}
H = \hbar\omega_0 |\Psi_0 \rangle\langle \Psi_0 | + \int \hbar\epsilon |\psi_\epsilon \rangle\langle 
\psi_\epsilon | + \int d\epsilon \left[ k(\omega_0, \epsilon ) |\Psi_0 \rangle\langle 
\psi_\epsilon |
+ k(\epsilon , \omega_0 ) |\psi_\epsilon \rangle\langle \Psi_0 |\right], \label{hammi}
\end{equation}
where by $\{\psi_\epsilon  \}$ we denote the continuum of eigenvectors outside the barrier. 
We then write the wavefunction in the form
\begin{equation}
|\Psi (t) \rangle = a (t) e^{-i\omega_0 t} |\Psi_0\rangle + \int d\epsilon b (\epsilon, t) e^{-i\epsilon t}
|\psi_\epsilon \rangle ,
\label{lapl}
\end{equation}
with $a(0) = 1$, $b (0) = 0$. Inserting this expression in the Schr\"odinger equation we get
an integro-differential equation for $a (t)$. The Laplace transform of this equation reads
\begin{equation}
{\cal L} [a] (s) = \frac{1}{s + {\cal L}[\chi ](s)},\label{laplace}
\end{equation}
where
\begin{equation}
\chi (t) = \frac{1}{\hbar ^2}\int d\epsilon |k (\epsilon ,\omega_0 )|^2e^{i (\omega_0 - \epsilon )t}.
\end{equation}
In general, the tunneling matrix element $k(\omega_0, \epsilon)$ 
depend on the energies $\epsilon$ and they are determined by
the overlap of the left and right wavefunctions under the barrier
\cite{cond}.
We notice that since typically the lifetime of the metastable states is much larger than the
oscillation period in the well (in other words the last term in the Hamiltonian is a perturbation),
 the states $\{|\epsilon \rangle \}$ which contribute effectively to tunneling
are located in a relatively small energy interval compared to the plasma oscillation frequency, 
therefore the shape of these states under the barrier
is approximately identical. We can then take
the tunneling matrix element as being a complex constant; but since we
will be interested exclusively in the outgoing component, the
relative phase between the wavefunction inside the well and that outside
will not play any role. We have confirmed this assumption also by 
expanding the initial wavefunction in terms of the WKB solution
of the washboard potential calculated in \cite{josephson}.
Therefore
we take $k = \hbar\sqrt{\Gamma /2\pi}$ real; 
we obtain ${\cal L}[\chi] (s) = \Gamma /2$= constant, which turns out to be
 the decay probability of the system. Indeed, the inverse Laplace transform of  Eq. (\ref{laplace}) gives 
precisely the classical exponential decay law
\begin{equation}
a (t) = e^{-\Gamma t/2}.
\end{equation}
The outgoing wavepacket becomes 
\begin{equation}
|\Psi_{out} (t) \rangle = -i\sqrt{\frac{\Gamma }{2\pi}}\int d\epsilon \frac{e^{(-\Gamma /2 + i \omega_0 )t}- e^{-i\epsilon t}}{\Gamma /2 + i (\omega_{0} -\epsilon )}
|\psi _\epsilon^{\pm} \rangle .
\label{q}
\end{equation}
To conclude this derivation, we find that with the identification $k = \hbar\sqrt{\Gamma /2\pi}$ the Hamiltonian
Eq. (\ref{hammi}) becomes a model Hamiltonian for decay in the continuum which can be solved exactly, with 
solution given by Eqs. (\ref{psii}) and (\ref{q}). A similar type of model Hamiltonian has been obtained in 
\cite{twopotential,kurizkikofman}.
One can show, using the properties of the 
Lorentz distribution, that
these wavefunctions are correctly normalized, as explained above. 

Let us now single out one component of the wave $\Psi_{out} (t)$, namely
\begin{equation}
\Psi^{\rightarrow}_{out} = i\sqrt{\frac{\Gamma }{2\pi}}\int d\epsilon \frac{e^{-i\epsilon t}}
{\Gamma /2 + i (\omega_{0} -\epsilon )}
|\psi _\epsilon^{+} \rangle .
\end{equation}
We first notice that the normalization of the total function $|\Psi_{out}\rangle$ is such that 
$\langle \Psi_{out}|\Psi_{out}\rangle = 1 - \exp (-\Gamma t)$, which reflects correctly the fact that
the probability of finding the particle outside comes from an exponential decay law, 
while that of $|\Psi^{\rightarrow}_{out}\rangle$ 
is such that $\langle \Psi^{\rightarrow}_{out}|\Psi^{\rightarrow}_{out}\rangle = 1$. In the following
we will see that $|\Psi^{\rightarrow}_{out} \rangle$ plays indeed a special role. 
To move on, we notice that part of the expression for the outgoing phase contains a term 
which decays exponentially on a time scale $\Gamma^{-1}$. These terms are associated with the 
fast components of the localized wavefunction which would escape first. 
Although a calculation that includes these terms is no doubt interesting, especially for the problem of
non-exponential decay rates \cite{deltapotential}, 
in what follows 
we will regard them as transient oscillatory effects whose presence will be difficult to assess experimentally 
anyway, and we will neglect their contribution.  
In the WKB approximation, far enough from the classical turning point, the eigenvalues 
$\{ |\psi _\epsilon \rangle \}$ have the form (up to a normalization factor and constant phase factors 
due to matching to the region left of the classical turning point) of incoming and outgoing scattering states
\begin{equation}
\psi^{\pm}_\epsilon (\gamma ) \approx \sqrt{\frac{e}{C\bar\Phi _0 V_{\epsilon}(\gamma )}}\exp\left[\pm i\frac{C}{e}
\int_{\gamma_0}^{\gamma}V_{\epsilon }(\varphi)d\varphi \right] ,\label{wkb}
\end{equation}
where 
\begin{equation}
V_{\epsilon}(\gamma ) = \frac{\hbar\omega_p}{\sqrt{2}e}\sqrt{\frac{2e\epsilon  + I \gamma }
{I_{c0}} + 1 - \cos\gamma }.
\end{equation}
The physical meaning of this voltage is that it corresponds to the (classical) energy accumulated 
on the capacitor when the phase difference across the junction is $\gamma$ and the initial energy
of the system is $\epsilon$; indeed, $CV^{2}_{\epsilon}(\gamma )/2 = \hbar\epsilon - U(\gamma )$. 
We now use the fact that for values of $\gamma$ outside the well and far enough from the classical turning
point the inequality $|\hbar\epsilon - \hbar\omega_{0}|\ll \hbar\omega_0 - U(\gamma )$ holds. 
We can therefore take $V_\epsilon (\gamma ) = V_{\omega_0}(\gamma )\stackrel{\rm not}{=}
V_{0}(\gamma )$ in the denominator of Eq. (\ref{wkb})
and approximate the exponent as
\begin{equation}
V_{\epsilon} (\gamma ) \approx V_{0}(\gamma )\left[1 + \frac{\epsilon - \omega_{0}}{2( 
\omega_{0}- \hbar^{-1}U( \gamma ))}\right].
\end{equation}
With these approximations, using Eqs. (\ref{q}) and (\ref{wkb}) we can performe the integral over 
the angular frequencies $\epsilon$; as a result, the contribution of the in-going scattering states in zero, 
while the out-going scattering states build up a wavepacket of the form 
\begin{equation} 
\Psi^{\rightarrow}_{out}(\gamma ,t) = \frac{{\cal N}}{\sqrt{V_{0}(\gamma )}}
\exp\left[\frac{i}{\hbar}\int_{\gamma_{0}}^{\gamma } 
C \tilde{V}(\varphi )\bar\Phi_0 d\varphi -
\left(i\omega_{0} + 
\frac{\Gamma}{2}\right)\left(t-\int_{\gamma_0}
^{\gamma}
\frac{\bar \Phi_{0}d\varphi }{V_{0}(\varphi)}\right)\right]
\Theta\left[t - \int_{\gamma_0}^{\gamma}
\frac{\bar \Phi_{0}d\varphi }{V_{0}(\varphi)}\right]. 
\label{out}
\end{equation}
Here ${\cal N}$ is a normalization factor which can be obtained through $\int_{-\infty}^{\infty} 
d(\bar\Phi_{0}\gamma )|\Psi^{\rightarrow}_{out}(\gamma ,t)|^2 = 1$
with the mention that we make a negligible error by extending the integral to $-\infty$, 
{\it i.e.} before the well region (where the actual values are exponentially small). The voltage
$\tilde{V}(\gamma )$ is defined as
$\tilde{V}(\gamma )= V_{0}(\gamma ) - \hbar\omega_0 /C V_{0}(\gamma )\approx V_{0}(\gamma )$.
It is interesting to see also what happens with the rest of the components of $|\Psi_{out}\rangle$.
Although they do contribute to the normalization as discussed before, they are decaying both in time and 
away from 
the barrier as $\exp [-\frac{\Gamma}{2}\int_{\gamma_0}^{\gamma}V_{0}^{-1}(\varphi)
\bar \Phi_{0}d\varphi ]$
which, as we will see below, would give far from the barier a factor of $\exp [-
\Gamma /2(t + \sqrt{2\gamma}/\omega_{p})]$. It is clear that these terms can be neglected starting
roughly from a time $\Gamma^{-1}$.   
The wavefunction Eq. (\ref{out}) contains all the information about the dynamical evolution of the state of the
circuit containing the Josephson junction, and it is the main result of this paper. 

In a typical experiment, 
the voltage across the junction is 
monitored by a voltmeter at room temperature. A fundamental issue is to find a microscopic mechanism for the junction-voltmeter interaction and 
a suitable theory of quantum measurement that would
model the collapse of the wavefunction; this is however beyond the scope of this paper. 
Still, Eqs. (\ref{out}) and (\ref{prob}) give a quite clear qualitative  
picture of what happens: the 
particle rolls down the washboard potential with a quasi-classical speed given by energy conservation
$CV^{2}_{0}(\gamma )/2 = \hbar\omega_0 - U(\gamma )$. Quantum mechanics enters in the picture through the tunneling rate; we expect the results
of the measurements 
to have a spread dermined by $\Gamma$.
One can assume that the measurement projects the outgoing state onto eigenvalues of the voltage
operator; therefore the  
probability of recording the value $V$ at the moment $t$ will be given by the standard
quantum mechanics recipe
\begin{equation}
P(V, t) = 
\frac{1}{2\pi\hbar}\vert \int_{-\infty}^{\infty} d(\bar\Phi_0\gamma )\Psi_{out}^{*}(\gamma ,t)\exp{(iVC\gamma /2e)} \vert ^2.
\label{prob}
\end{equation}
 As an example, had the outside the well  potential $U$ been
zero, we would have gotten for the charge $CV$, by performing the integration in Eq. (\ref{prob}), a 
standard Cauchy-Breit-Wigner distribution centered around $CV_0$ and full width at half maximum 
$\Gamma\hbar /V_0$
\begin{equation}
P(CV)= \frac{1}{\pi}\frac{\Gamma\hbar}{2 V_0}\left[ (CV-CV_0)^2 + 
\left( \frac{\Gamma \hbar}{2V_0}\right)^2 \right]^{-1}.
\end{equation} 



\begin{figure}[htb]
\includegraphics[width=84truemm]{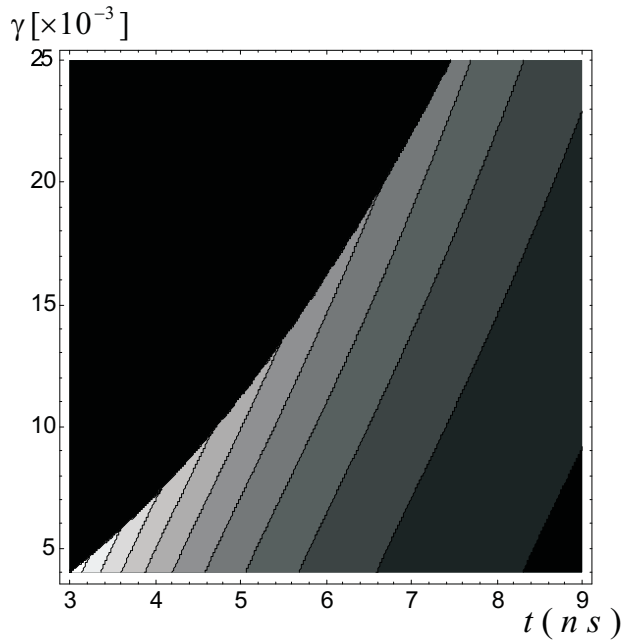}
\caption{Contourplot $(\gamma, t)$ of the modulus of the running-phase wavefunction Eq. (\ref{super}) for $\omega_p = 60 \Gamma = 30$ GHz.}
\label{wf}
\end{figure}

Considering again the case of a junction with a washboard potential $U(\gamma )$, we notice that
a good approximation is $U (\gamma ) \approx E_{J}\gamma $. This comes from the fact that switching is 
typically observed at values of the bias current close to the critical current of the junction, as well
as from the observation that for times larger than $\Gamma ^{-1}$ the wavepacket is concentrated at
large values of $\gamma \gg 1$, in which case the $\cos\gamma$ term in the potential is negligible. 
In other words, the particle gets soon so fast that the "speed bumps" created by the Josephson 
effect are not slowing it down significantly. This can be checked {\it a posteriori}. A first observation
is that the relevant quantity for the dynamics of the center of the wavepacket is the argument of the
$\Theta $ function; the condition that this argument vanishes sets the maximum value
of $|\Psi_{out}|^2$ and gives a phase $\omega_{p}t/2\gg 1$ for $t$ larger than $\Gamma^{-1}$. 
A legitimate concern is whether the wavefunction does not spread faster than it moves downwards. This 
is not the 
case, as we will see below: the spread of the wavefunction increases linearly with time, while the 
average coordinate (phase) is advancing as $t^2$. With these observations, the normalization constant
can be calculated, and the outgoing wavepacket becomes
\begin{equation}
\psi^{\rightarrow}_{out} (\gamma ,t) = \sqrt{\frac{\Gamma}{\sqrt{2\gamma}\bar\Phi_{0}\omega_{p}}}\exp\left[i\frac{\hbar\omega_{p}
\gamma^{3/2}}{6\sqrt{2}E_{c}} - \left(i\frac{\omega_0}{\omega_p} + \frac{\Gamma}{2\omega_p}\right)\left(
t\omega_{p}-\sqrt{2\gamma}\right)\right]\Theta\left(t\omega_p - \sqrt{2\gamma}\right).\label{super}
\end{equation}
A plot of the wavefunction is given in Fig. \ref{wf}. 

The average phase (flux) corresponding to this wavepacket can be obtained
\begin{equation}
\langle \gamma \rangle (t)= \omega_{p}^2(t^{2}/2 - \Gamma ^{-1}t + \Gamma^{-2}),\label{av}
\end{equation}
and we notice that the dominant term is quadratic in t. 
The spread of the flux variable is given by (we keep only the dominant term here)
\begin{equation}
\sqrt{\langle \gamma^2\rangle (t) - \langle \gamma \rangle^2(t)} = \omega_{p}^{2}\Gamma^{-1}t.
\end{equation}
To get the average voltage we can use Ehrenfest theorem; we obtain
\begin{equation}
\langle V(t) \rangle = \bar\Phi_{0}\omega_{p}^{2}(t -\Gamma^{-1}).  \label{avv}
\end{equation}  
The dominant term for the voltage is linear in time and satisfies the classical energy
conservation $C\langle V\rangle ^2(t) /2 = \hbar \omega_{0} - U(\langle \gamma \rangle (t))\approx I_{0}\langle
\gamma \rangle (t)$.  

\begin{figure}[htb]
\includegraphics[width=84truemm]{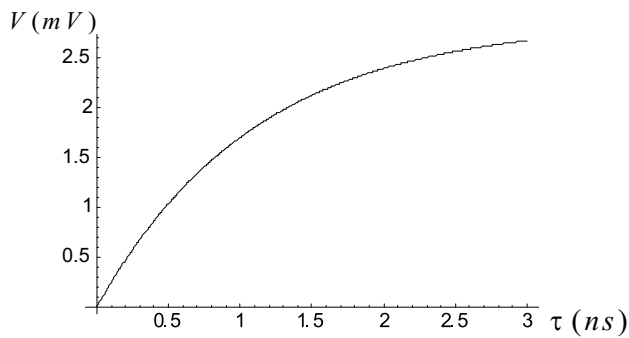}
\caption{The average voltage as a function of the time $\tau$ for $\omega_p = 30$ GHz, $\Gamma = 1$ ns 
 and $t = 10$ ns.}
\label{prediction}
\end{figure}

Let us now analyze what happens in typical switching current experiments, as they are done now in the
context of superconducting qubits \cite{qubit}: the bias current of the junction is increased
fast to a value that allows tunneling, it is kept there for a time $0 < \tau < \Gamma^{-1}$, then it is lowered
to a value that suppresses tunneling. This value has to be large enough so that the experimentalist can get a 
reliable reading of voltage on the quasiparticle branch if the junction has switched; in practice, it can still satisfy
$I\approx I_0$. Although the change 
of the bias current has a major effect with respect to tunneling through the barrier, where the 
the tunneling rate decreases exponentially with the height of the barrier, from the point of view of the
structure of the running-phase state it amounts only to a modification of the parameter $\Gamma$.  
Finally,  
the current is put to zero and, after waiting long enough for retrapping to occur, the whole cycle can be 
repeated. In our model, the essential physics is that after the time $\tau$, the tunneling matrix element $t$ is zero,
therefore the system evolves only under the action of $H_0$. The wavefunction is "cut" into two separate pieces,
one which is (almost) the bound state $\Psi_0$ inside the well, the other being the wavepacket in the continuum
which evolves as
\begin{equation}
|\Psi_{out} (t) \rangle = -i\sqrt{\frac{\Gamma }{2\pi}}\int d\epsilon \frac{e^{(-\Gamma /2 + i \omega_0 )\tau }
e^{-i\omega_{j} (t-\tau )} 
- e^{-i\epsilon t}}{\Gamma /2 + i (\omega_{0} -\epsilon )}
|\psi _\epsilon^{\pm}\rangle ,
\end{equation}
with normalization $\langle \Psi_{out} (t)|\Psi_{out} (t) \rangle = 1- e^{-\Gamma \tau}$. Now, for $t-\tau >
\Gamma^{-1}$ we can see
that the outgoing function consists of two consecutive (separated by the time $\tau$) and dephased (with $\omega_0 \tau $)
outgoing wavepackets with the structure of $|\Psi^{\rightarrow}_{out} (t)\rangle$
which propagate at the same speed across the phase coordinate $\gamma$. The second wavepacket, which has a probability amplitude
smaller by a factor of $e^{-\Gamma /2\tau}$, results from the
waves localized near the barrier during the time $\tau$ when tunneling was in progress. 
After integration over energy, we get
\begin{equation}
|\Psi_{out} (t) \rangle = e^{-(\Gamma /2 + i\omega_0)\tau }|\Psi^{\rightarrow}_{out} (t -\tau )\rangle
- |\Psi^{\rightarrow}_{out} (t)\rangle ,\label{xx}
\end{equation}
where $|\Psi^{\rightarrow}_{out} (t)\rangle$ is given by Eqs. (\ref{out}) and (\ref{super}). To check that 
the normalization 
$\langle \Psi_{out} (t)|\Psi_{out} (t) \rangle = 1- e^{-\Gamma \tau}$ remains valid, we notice that 
in the region of overlap of the two wavepackets, which coincides with the domain where 
$|\Psi^{\rightarrow}_{out}(t-\tau )\rangle$
is finite $t-\tau > \int_{\gamma_0}^{\gamma}\bar \Phi_{0} V_{0}^{-1}(\varphi)d\varphi $, there exists a very 
simple relation between them: 
$|\Psi^{\rightarrow}_{out} (t) \rangle = \exp[-i\omega_0\tau -\Gamma \tau /2]|\Psi^{\rightarrow}_{out} (t-\tau )\rangle$.
Using this property and the previous expressions Eqs. (\ref{av}) and (\ref{avv}), we can calculate the average
phase and voltage on the state Eq. (\ref{xx})
\begin{equation}
\langle \gamma \rangle (t) = \frac{1}{2}\omega_{p}^{2}[1- e^{-\Gamma \tau}]t^{2} - 
e^{-\Gamma \tau}\omega_{p}^{2}\tau t - \frac{1}{2}e^{-\Gamma \tau}\omega_{p}^{2}, 
\end{equation}
and 
\begin{equation}
\langle V \rangle (t) = \bar\Phi_{0}\omega_{p}^{2}t\left(1- e^{-\Gamma \tau }\right)
-\bar\Phi_{0}\omega_{p}^{2}e^{-\Gamma \tau}\tau .
\end{equation}
In Fig. \ref{prediction} we present a plot of the average voltage as a function of the time $\tau$.
We see that for values of $\tau$ of the same order or larger than the lifetime $\Gamma^{-1}$
the average voltage at t flattens, reflecting the fact that the junction has switched, as in the case
of Eq. (\ref{avv}).  

For designing an experiment to test these predictions, several remarks should be made. 
In the case of real junctions, the Josephson energy and the plasma frequency can be reduced 
by using a SQUID configuration and by adding capacitors in parallel with the junctions.
This makes the time evolution of the switching state slower and therefore easier to detect.
An important limitation on time comes from the fact that as soon as the voltage reaches the quasiparticle
branch (at twice the value of the gap) our analysis is not valid. The other limitation is technological: 
even with a good dilution refrigerator,
thermalizing the junction is very difficult at low temperatures. With a good high-power refrigerator
with base temperature of about 5 mK, we assume an optimistic value of 10 mK for the effective temperature
of the electrons. This temperature corresponds to a crossover angular frequency of 8.66 GHz between the MQT
and the thermal
activation transition. 
A plasma frequency of $\omega_p = 30$ GHz (zero bias current) will thus keep us safely in the MQT regime when 
the current is raised up to about half a percent close to the critical current, according to the formula that 
gives the plasma oscillation frequency at a finite bias current \cite{tinkham,switch}. For Nb, with gap of 1.4 meV, this 
corresponds to a time of approximately 10 ns, as given by Eq. (\ref{avv}). A voltage increase on this timescale 
can be detected with standard experimental techniques
Suppose now that we choose to work at currents about 5\% less than the critical current.  
We still have to satisfy the condition $t>\Gamma^{-1}$; an inspection of the formula that gives 
the tunneling rate for underdamped junctions (see {\it e.g.} \cite{switch}) shows that switching rates of about 500 MHz and more 
(with the restriction $\Gamma \ll \omega_p$) can be achieved for $E_J /\omega_p$ of the order of 30, values which can be obtained easily with 
large junctions.

G.~S.~P. was supported by an EU  Marie Curie Fellowship (HPMF-CT-2002-01893); this work is also
part of the SQUBIT-2 project (IST-1999-10673), the 
Academy of Finland TULE No.7205476, and the 
Center of Excellence in Condensed Matter and Nuclear Physics at 
the University of Jyv\"askyl\"a.

\end{document}